\documentclass[prb,twocolumn,amsmath,showpacs]{revtex4-1}
\usepackage{hyperref,graphicx}

\def\subfiga{(i)}
\def\subfigb{(ii)}
\def\subfigc{(iii)}

\begin{document}

\title{Influence of spin dynamics of defects on weak localization in paramagnetic two-dimensional metals}

\author{Oleksiy Kashuba}
\affiliation{Institut f\"ur Theoretische Physik und Astrophysik, Universit\"at W\"urzburg, Am Hubland, D-97074 W\"urzburg, Germany}
\affiliation{Institute of Theoretical Physics, Technische Universit\"at Dresden, D-01062 Dresden, Germany}
\email{okashuba@physik.uni-wuerzburg.de}

\author{Leonid I.\ Glazman}
\affiliation{Department of Physics, Yale University, New Haven, Connecticut 06520, USA}

\author{Vladimir I.\ Fal'ko}
\affiliation{National Graphene Institute, University of Manchester, Booth Street East, Manchester M13 9PL, United Kingdom}


\date{\today}

\begin{abstract}
Spin-flip scattering of charge carriers in metals with magnetic defects leads to the low-temperature saturation of the decoherence time $\tau_\varphi$ of electrons at a value comparable to their spin relaxation time $\tau_s$.
In two-dimensional (2D) conductors such a saturation can be lifted by an in-plane magnetic field $B_\parallel$, which polarizes spins of scatterers without affecting the orbital motion of free carriers.
Here, we show that in 2D conductors with substantially different values of the $g$ factors of electrons ($g_e$) and magnetic defects ($g_i$), the decoherence time $\tau_\varphi(B_\parallel)$  (reflected by the curvature of magnetoconductance) displays an anomaly: It first gets shorter, decaying on the scale $B_\parallel\sim \hbar/|g_i-g_e|\mu_B \tau_s$, before becoming longer at higher values of $B_\parallel$.
\end{abstract}

\maketitle

The electron interference results in a quantum correction to the Drude conductivity and a positive magnetoconductivity (MC) of disordered metals.~\cite{Altshuler1980}
In particular, the constructive interference of electron waves propagating in time-reversed fashion along closed diffusive loops in two-dimensional (2D) conductors brings about a logarithmically divergent weak localization (WL) correction. 
In the absence of external magnetic flux piercing the electron trajectories, this divergence of WL correction is cut off by the electron decoherence time $\tau_\varphi$. 
Application of the flux breaks the time-reversal symmetry, thus further diminishing the WL correction and leading to the low-temperature MC, $\sigma (B_{\perp})$, where $B_\perp$ is the  magnetic field component perpendicular to the plane of a 2D sample. 
The MC curvature, $\kappa \equiv \left. \frac{\partial ^2 \sigma}{\partial B_{\perp}^2} \right|_{B_{\perp}=0} \propto \tau_\varphi^2$ gives a measure for the electron coherence time in 2D conductors: doped semiconductor quantum wells, charge accumulation layers near semiconductor interfaces, thin metallic films, or atomically thin 2D crystals such as graphene and transition-metal dichalcogenides monolayers.

\begin{figure}
\centering
\includegraphics[scale=.67]{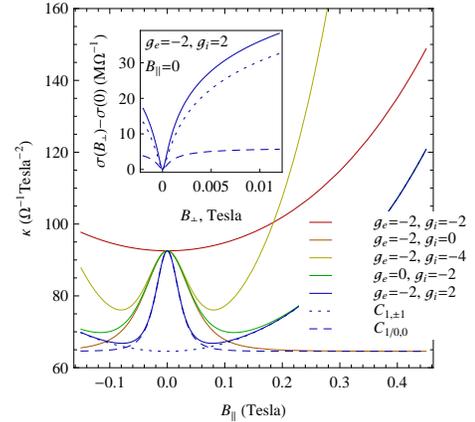}
\caption{Dependence of the $B_\perp=0$ MC curvature, $\kappa \propto \tau_\varphi^2$ on the in-plane magnetic field [see Eq.~\eqref{eq:csmf}], calculated for spin-$\frac12$ impurities with various $g$ factors: red, yellow, and blue solid lines.
Dotted and dashed blue lines demonstrate the magnetic field dependence of Cooperons without [$C_{1,1}(B_{\parallel}) \!+\! C_{1,-1}(B_{\parallel})$] and with [$- C_{0,0}(B_{\parallel}) \!+\! C_{1,0}(B_{\parallel}) \!+\! C_{1,1}(0) \!+\! C_{1,-1}(0)$] spin-exchange processes for the case $g_i=2$.
The inset shows the respective contributions to the full MC curve in the absence of an in-plane magnetic field [see Eq.~\eqref{eq:mc}] and text for details. 
Here, $T=0.5\,\text{K}$, $D=100\,\text{cm}^{2}/\text{s}$, and $\tau_s=0.1\,\text{ns}$.}
\label{fig:curvatures}
\end{figure}

The two leading decoherence processes at low temperatures stem from the inelastic scattering of electrons off each other and off magnetic impurities. The electron-electron scattering results~\cite{Altshuler1985,Aleiner1999} in the linear temperature dependence of the decoherence rate, $\tau_T^{-1} = \frac{kT}{\hbar}\frac{e^2/h}{\sigma}\ln\frac{\sigma}{2e^2/h}$, with $\tau_{T\to 0}^{-1}\to 0$.
In contrast, the contact exchange interaction with paramagnetic defects results in an apparent%
~\footnote{%
In fact, this contribution to the decoherence rate also vanishes at $T\to 0$, but the characteristic temperature scale is defined by the Kondo temperature~\cite{Glazman2003} and in many cases is extremely low.%
} %
low-temperature saturation~\cite{Larkin1980,Stone1987,Stone1989,Falko1991,Chandrasekhar1990,Geim1990,Levy1991,Birge2000,Birge2002,Birge2003} of the electron decoherence rate at $\tau_\varphi^{-1}(T\to0)\sim\tau_s^{-1}$.
The rate $\tau_s^{-1}$ characterizes the electron spin relaxation due to the spin flips in the course of electron scattering off randomly oriented magnetic moments of impurities.

It is common knowledge~\cite{Glazman2003,Falko1992,Amaral1990} that electron spin relaxation may be suppressed and $\tau_\varphi$ extended by the polarization of magnetic impurities.
In 2D conductors, this can be achieved by using an in-plane magnetic field, $B_\parallel$, which polarizes the spins of the impurities but does not create any flux through the electron orbits.
Measurements of $\kappa$ in various low-dimensional materials~\cite{Geim1990,Levy1991,Birge2000,Birge2002,Birge2003} have shown a gradual increase of $\tau_\varphi (B_\parallel)$ associated with the spin polarization at $g_i\mu_B B_\parallel S\gtrsim T$ (here, $g_i$ and $S$ are respectively the $g$ factor and spin of a magnetic impurity; $\mu_B=|e|\hbar/2mc$ is the Bohr magneton).

Here, we show that in some 2D materials the  dependence of decoherence time on the in-plane magnetic field, $\tau_\varphi(B_\parallel)$, may be nonmonotonic: The naively expected polarization-induced increase of $\tau_\varphi$ with $B_\parallel$ is preceded by its decrease at weak fields (see Fig.~\ref{fig:curvatures}).
This acceleration of decoherence comes from the precession dynamics of localized magnetic moments and requires the $g$ factors of the impurities ($g_i$) and electrons ($g_e$) to differ from each other.
To mention, if $g_i=g_e$, then the local moments are static in the frame rotating together with the precessing electron spins, and in this case the $\tau_\varphi(B_\parallel)$ dependence remains monotonic, being caused solely by the impurity spin polarization.
For $g_i\neq g_e$, electrons witness the landscape of magnetic moments that varies in time with the frequency
\begin{equation}
\Omega_B = \frac{(g_e-g_i)\mu_B B_\parallel}{\hbar}.
\label{eq:omb}
\end{equation}
This temporal variation shortens $\tau_\varphi$, if $\Omega_B \tau_s\gtrsim 1$.
The latter condition is satisfied already at nonpolarizing fields, assuming that $|1-g_e/g_i|kT\tau_s/\hbar S\gg 1$.

\begin{figure}
\centering
\includegraphics[scale=.55]{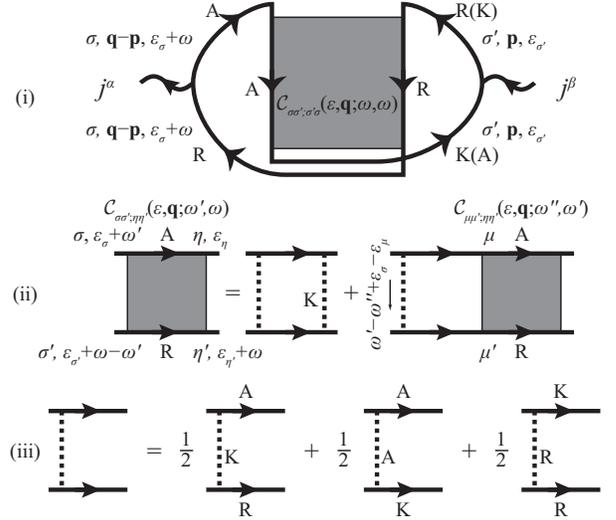}
\caption{Disorder perturbation theory diagrams for the WL correction to conductivity.
\subfiga\ WL correction related to the Cooperon $\mathcal{C}_{\sigma\sigma';\eta\eta'}(\varepsilon,\mathbf{q};\omega',\omega)$.
Bold dots stand for the current operators and bold lines are the disorder-averaged Keldysh functions.
\subfigb\ Bethe-Salpeter equation for the Cooperon.
\subfigc\ Combination of components of Keldysh functions involved in the kernel of the Cooperon equation.
The first diagram in the right part of the equation contains all types of scattering described in Fig.~\ref{fig:scattline}(a--d), while the second and third contain the spin-exchange processes (c,d) only.}
\label{fig:kubobethesalp}
\end{figure}

Polarization of the magnetic impurities renders them ineffective in the electron phase relaxation, thus leading to a strong increase of the magnetoconductance signal.~\cite{Geim1990,Levy1991,Birge2000,Birge2002,Birge2003,Glazman2003,Falko1992,Amaral1990}
Contrary to that, the effect of spin dynamics is quite subtle.
We find the limitation on the magnitude of the corresponding change in the magnetoconductance, evaluate analytically the dependence of the $\tau_\varphi$ on $B_\parallel$, and relate it to the basic parameters of the itinerant electrons and magnetic impurities. 

To analyze the influence of spin-flip scattering in a conductor on the WL effect, we consider an electron wave propagating along a closed-loop trajectory, scattering from disorder, 
$V(\mathbf{r}) =  \sum_{\mathbf{r}_{i}} U \delta(\mathbf{r}-\mathbf{r}_{i}) + \sum_{\mathbf{r}_{j}}  J\,\mathbf{S}_{j}\cdot \mathbf{\sigma}\,\delta(\mathbf{r}-\mathbf{r}_{j})$.
The Zeeman terms for electrons and impurities are $ -g_{e}\mu_{B}B_{\parallel}\sigma_z$ and $-g_i\mu_B B_\parallel(\mathbf{S}_{j})_{z}$, respectively (the $z$ axis is chosen along the in-plane magnetic field), and $\sigma$ is the electron spin operator acting on the $\pm^1\!/_2$ spin states quantized along the $z$ axis.

To quantify the $\tau_\varphi (B_\parallel)$ dependence, we express the WL correction $\delta\sigma$ to conductivity~\cite{Larkin1980} in terms of two-electron propagators, ``Cooperons'' $\mathcal{C}_{\sigma\sigma';\eta\eta'}(\varepsilon,\mathbf{q};\omega',\omega)$ [see Fig.~\ref{fig:kubobethesalp}\subfiga]:
\begin{multline}
\delta\sigma = \frac{e^2}{2\pi h} \left[ C_{0,0} - C_{1,0} - C_{1,1} -C_{1,-1} \right], \\
C_{S,M} = \zeta_{S,M;\sigma\sigma'} \hat{C} \zeta_{S,M;\sigma\sigma'}^\mathsf{T}, \\
\hat{C}  = -\gamma v_{F}^{2}\tau^{3}\int d\varepsilon d^{2}\mathbf{q}
n_{F}'(\varepsilon_{\sigma'})
\hat{\mathcal{C}}(\varepsilon,\mathbf{q}).
\label{eq:WLconductivity}
\end{multline}
Here, Clebsch-Gordan coefficients $\zeta_{S,M;\sigma\sigma'}=\langle S,M|^{1}\!/_{2},\sigma;^{1}\!\!/_{2},\sigma'\rangle$ select from the Cooperon matrix $\hat{\mathcal{C}}(\varepsilon,\mathbf{q}) \equiv \int d\omega\,\mathcal{C}_{\sigma\sigma';\eta\eta'}(\varepsilon,\mathbf{q};\omega,\omega)$ the singlet ($S=0$) and triplet ($S=1$, $M=-1,0,1$) components defined in terms of the total spin carried by the two-electron propagator and its projection onto the external magnetic field $\mathbf{B}_\parallel$.
Also,
\begin{equation}
\varepsilon_{\sigma}=\varepsilon - \sigma g_{i}\mu_{B}B_{\parallel},
\label{eq:epm}
\end{equation}
and $n_{F}'(\varepsilon)\equiv\partial n_F(\varepsilon) /\partial\varepsilon$ is a derivative of the Fermi distribution function.%
~\footnote{%
Note $g_{i}$ (not $g_{e}$) in this equation, which is because the Green's function of an electron with spin $\sigma$ has energy $\varepsilon_{\sigma}$.
The energy of the electron changes only at the spin flip, always by the value $\pm g_{i}\mu_{B}B_{\parallel}$ dependent on the sign of the transmitted spin $\sigma-\sigma'$.
Therefore, in the final spin state $\sigma'$, the electron will have the energy $\varepsilon_{\sigma}\pm g_{i}\mu_{B}B_{\parallel}=\varepsilon_{\sigma'}=\varepsilon_{-\sigma}$.%
}

\begin{figure}
\centering
\includegraphics[scale=.55]{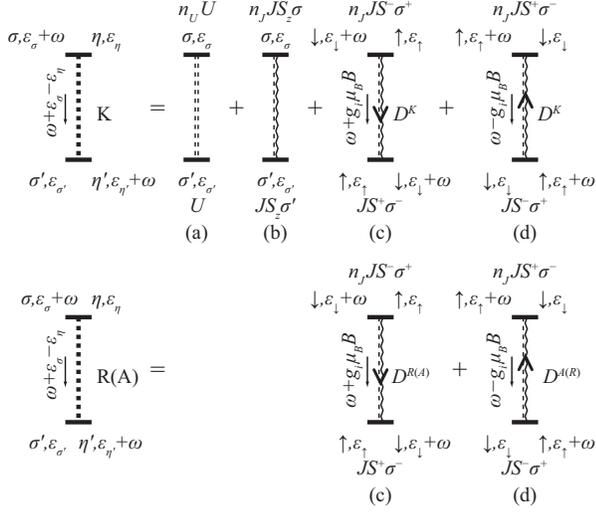}
\caption{Disorder correlation functions.
First line: (a) Keldysh part of spin-independent scattering, (b) elastic spin-dependent scattering without spin-flip;
(c,d) spin-flip scattering, with the energy transfer in the spin-flip process shown along the dashed line.
Second line: (c,d) Retarded/advanced correlator components.
All (c,d) lines being implemented in Fig.~\ref{fig:kubobethesalp}\subfigc\ describe the spin exchange between two electrons and appear only in the Cooperons $C_{1/0,0}$.
}
\label{fig:scattline}
\end{figure}

The diagrammatic form of the Bethe-Salpeter equation for the Cooperon matrix  $\mathcal{C}$ is shown in Fig.~\ref{fig:kubobethesalp}\subfigb.
Its important element is the disorder correlation function represented by the dashed lines in Fig.~\ref{fig:scattline}, which is assumed to be short ranged and includes the following elements:

\noindent
(a) Correlator of spinless disorder,
$n_{U}U^{2} \delta(\omega)\delta(\mathbf{r}-\mathbf{r}')$,
where $n_{U}$ is the density of the pointlike potential scatterers.

\noindent
(b) Correlator of the $z$-spin components of local magnetic moments that characterizes spin-dependent scattering of electrons without spin flip,
$n_{J}J^{2} \langle \mathrm{T}_{K} S_{z}(t)S_{z}(t')\rangle  \delta (\mathbf{r}-\mathbf{r}')$.
Here, $n_{J}$ is the density of the magnetic defects of spin $S$.
The spin correlator $\langle \mathrm{T}_{K} S_{z}(t)S_{z}(t')\rangle$ is independent on the positions $t$ and $t'$ on the Keldysh contour, hence, it has only a Keldysh component with the Fourier transform
$n_{J}J^{2}\langle S_{z}^{2}\rangle\delta(\omega) \delta (\mathbf{r}-\mathbf{r}')$.

\noindent
(c,d) Spin correlators 
$n_{J}J^{2} \langle \mathrm{T}_{K} S_{+}(t)S_{-}(t')\rangle  \delta (\mathbf{r}-\mathbf{r}')$,
where $D(t,t')=\langle \mathrm{T}_{K} S_{+}(t)S_{-}(t')\rangle$ is mapped from the Keldysh time contour onto the matrix Keldysh space with components
\begin{equation}
\begin{split}
D^{R/A}(\omega)&=2i\langle S_{z}\rangle (\omega-g_{i}\mu_{B}B\pm i0)^{-1},
\\
D^{K}(\omega)&=4\pi [S(S+1)-\langle S_{z}^{2}\rangle]\delta(\omega-g_{i}\mu_{B}B).
\end{split}
\end{equation}
Here, $\langle S_{z}^{n}\rangle = [Z(a)]^{-1}\partial^{n}_{a}Z(a)$, $a=g_{i}\mu_{B}B_{\parallel}/kT$, and $Z(a)=\sum_{S_{z}=-S}^{S}e^{aS_{z}}$ is the partition function for a paramagnetic scatterer.

\begin{figure}
\centering
\includegraphics[scale=.55]{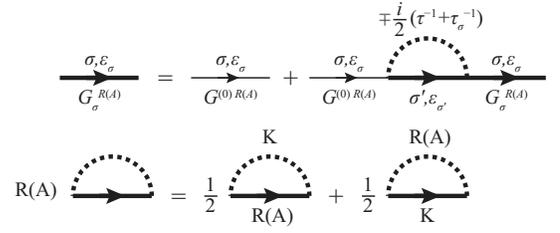}
\caption{First line: Diagrammatic representation of Dyson's equation for the single electron Green's functions calculated in the main order in $\hbar/vp_{F}\tau \ll 1$.
Second line: Keldysh structure of the self-energy of the retarded/advanced Green's functions.}
\label{fig:greenfunc}
\end{figure}

The thick solid lines in Fig.~\ref{fig:kubobethesalp} stand for disorder-averaged electron Green's functions $G_{\sigma}$, obtained from the solution of the Dyson equation shown in Fig.~\ref{fig:greenfunc},
\begin{equation}
\begin{split}
G^{R/A}_{\sigma} &= \!\left(\!\varepsilon_{\sigma} \!-\! v_{F}\xi_{\mathbf{p}} \!+\! g_{e}\mu_{B}B_{\parallel}\sigma
\pm \frac{i}{2}(\tau^{-1}\!+\!\tau_{\sigma}^{-1})\!\right)^{-1}\!,
\\
G^{K}_{\sigma} &= \bigl[(1-2n_{F}(\varepsilon_{\sigma})\bigr] \bigl(G^{R}_{\sigma}-G^{A}_{\sigma}\bigr), \\
\end{split}
\end{equation}
where $\tau=1/2\pi\gamma n_{U}U^{2}$ is the mean free time, $\tau_s=1/2\pi\gamma n_{J}J^{2}S(S+1)$ is the spin relaxation time, $\gamma$ is the electron density of states,
\begin{equation}
\begin{split}
\tau_{\sigma}^{-1} = \left[1 - 2\sigma M_{1} \bigl(1-2n_{F}(\varepsilon_{-\sigma})\bigr)\right]\tau_{s}^{-1},
\\
M_{n}=\langle S_{z}^{n}\rangle/S(S+1); \quad \xi_{\mathbf{p}}\approx v_{F}(|\mathbf{p}|-p_{F}),
\label{eq:rse}
\end{split}
\end{equation} 
and $\sigma = \pm^1/_2$ is the electron's spin projection on the direction of the in-plane magnetic field.

The spin structure of the Cooperons $C_{1,\pm1}$ allows only for the (a) and (b) contributions to the dashed line in the bottom row in Fig.~\ref{fig:kubobethesalp}\subfigc, forbidding the spin exchange (c,d), and thus securing $\mathcal{C}_{\sigma\sigma';\sigma\sigma'}(\varepsilon,\mathbf{q};\omega',\omega)\propto\delta(\omega')$.
The kernel of the Bethe-Salpeter equation for Cooperons $C_{1/0,0}$ includes spin-exchange contributions (c,d).
Summing up all three possible combinations of Keldysh function components $G^{A}D^{K}G^{R}$, $G^{A}D^{A}G^{K}$, and $G^{K}D^{R}G^{R}$ in Fig.~\ref{fig:kubobethesalp}\subfigc, where the frequency argument of the spin correlator is $D^{R/K/A}\equiv D^{R/K/A}(\omega+g_{i}\mu_{B}B)$, we get
\begin{multline*}
\frac{n_{J}J^{2}}{2}\Bigl( D^{K} + [1-2n_{F}(\varepsilon_{-}+\omega)](D^{A}-D^{R}) \Bigr)\\
=2\pi n_{J}J^{2}
\Bigl(
S(S+1) \!-\! \langle S_{z}^{2}\rangle \!-\! [1 \!-\! 2n_{F}(\varepsilon_{-})]\langle S_{z}\rangle
\Bigr)
\delta(\omega).
\end{multline*}
The frequency dependence of this kernel enforces $\hat{\mathcal{C}}_{\sigma\sigma';-\sigma,-\sigma'}(\varepsilon,\mathbf{q};\omega',\omega)\propto\delta(\omega')$, i.e.\ the energy transferred through the impurity spin correlator can be only $\pm g_{i}\mu_{B}B_{\parallel}$, where the sign depends on whether the spin transferred to the defect is $+1$, or $-1$.
After taking this into account, the equation for the Cooperon takes the form 
\begin{multline}
\left(
D\left(-i\hbar\partial_{\mathbf{r}}-\frac{2e}{c}\mathbf{A}_{\perp}\right)^{2}+\hat{R}
\right) \hat{\mathcal{C}}(\varepsilon,\mathbf{r})=\frac{1}{2\pi\gamma\tau^{2}}\delta(\mathbf{r});  \\
R_{\sigma\sigma';\eta\eta'} = \delta_{\sigma\eta}\delta_{\sigma'\eta'} \Bigl((\tau_{\sigma}^{-1}+\tau_{\sigma'}^{-1})/2 + \tau_{T}^{-1}\Bigr)
+  
\\
+ \delta_{\sigma+\sigma',\eta+\eta'}\Bigl(
- 4\sigma\sigma' \tau_{s}^{-1} M_{2}
+ |\sigma-\eta|\tau_{\sigma}^{-1} +\\
+ i(\sigma-\sigma')\Omega_B \Bigr),
\label{eq:BSe}
\end{multline}
where $D=v_{F}^{2}\tau/2$ is the diffusion coefficient, $\hat{\mathcal{C}}(\varepsilon,\mathbf{r})$ is the Fourier transform of $\hat{\mathcal{C}}(\varepsilon,\mathbf{q})$ and $\mathbf{A}_{\perp} = (0,B_{\perp}x)$ is the vector potential of the perpendicular magnetic field (note that $B_{\perp}\ll B_{\parallel}$).

Diagonalization of a matrix $\hat{R}$ produces Cooperons $C_{1,\pm 1}$ decoupled from all other Cooperon components and having decay rates
\begin{equation}
\tau^{-1}_{1,\pm1} = \left[1 - M_{2} \mp M_{1} \bigl(1 - 2 n_{F}(\varepsilon_{\mp}) \bigr) \right]\tau_{s}^{-1} + \tau_T^{-1}.
\label{eq:r1pm}
\end{equation}
Here, $\varepsilon_{\pm}$ [see Eq.~\eqref{eq:epm}] accounts for the energy transfer to an impurity in the process of spin-flip scattering.
The Cooperon components $C_{0,0}$ and $C_{1,0}$ are coupled with each other by spin-flip processes.
The coupling generates combined modes decaying with the rates
\begin{equation}
\begin{split}
\tau^{-1}_{0,\pm} =
\left[1 + M_{2} - M_{1} \bigl( n_{F}(\varepsilon_{+}) - n_{F}(\varepsilon_{-}) \bigr) \right] \tau_{s}^{-1}
\\
\pm \sqrt{\tau^{-1}_{1,+1} \tau^{-1}_{1,-1}-\Omega_B^2} +\tau_T^{-1}.
\label{eq:r0pm}
\end{split}
\end{equation}
Note that at $B_{\parallel}=0$ the average values $\langle S_{z}\rangle=0$ and $\langle S_{z}^{2} \rangle=\frac13 S(S+1)$, so that $\tau^{-1}_{1,\pm1}=\tau^{-1}_{0,-}=\frac23 \tau^{-1}_{s}+\tau_T^{-1}$ and $\tau^{-1}_{0,+}=2\tau^{-1}_{s}+\tau_T^{-1}$, in agreement with earlier theories.~\cite{Larkin1980,Stone1987,Stone1989,Falko1991,Glazman2003}

Relaxation rates $\tau^{-1}_{1,\pm 1}$ and $\tau^{-1}_{0,\pm}$ vary with $B_\parallel$ over two parametrically different field scales.
For $\tau^{-1}_{1,\pm 1}$, the scale is determined by the polarization of impurity spins.
The polarization takes place at $g_i\mu_B B_\parallel S \gg kT$, and Eq.~\eqref{eq:r1pm} then yields $\tau^{-1}_{1,\pm 1}-\tau_T^{-1}\ll\tau_s^{-1}$.
On the contrary, the field dependence of $\tau^{-1}_{0,\pm}$ is defined by the electron and impurity spin dynamics.
Under the condition $|1-g_e/g_i|kT\tau_s/\hbar S\gg 1$, the corresponding field scale is much smaller.
Neglecting the spin polarization, we may simplify Eq.~\eqref{eq:r0pm} to:
\begin{equation}
\tau^{-1}_{0,\pm} = \frac{4}{3\tau_{s}}
\pm \sqrt{\left(\frac{2}{3\tau_{s}}\right)^{2}-\Omega_B^{2}} +\tau_T^{-1}.
\label{eq:r0pms}
\end{equation}
As expected, the effect of the magnetic field depends on $\Omega_B$, the \emph{difference} between the precession frequencies of the impurity and electron spins [see Eq.~\eqref{eq:omb}].
The effect is absent if the corresponding $g$ factors are identical.

When substituted in Eqs.~\eqref{eq:WLconductivity}, the four Cooperon modes obtained using Eq.~\eqref{eq:BSe} yield the WL correction to the conductivity at $B_\perp=0$ (the first term in square brackets comes from $C_{1,\pm1}$ and the second from $C_{1/0,0}$),
\begin{equation}
\begin{split}
\delta\sigma \!=\! \frac{e^{2}}{2\pi h} \! \int d\varepsilon \! \sum_{\alpha=\pm} \!
n_{F}'(\varepsilon_{\alpha})
\!\left[
\ln\frac{\tau_{1,\alpha}}{\tau_\mathrm{min}}
+
A_{\alpha}\ln\frac{\tau_{0,-}}{\tau_{0,+}}
\right],
\\
A_{\pm} \!=\! (\tau^{-1}_{1,\pm1} \!-\! \tau_{T}^{-1})/(\tau_{0,+}^{-1} \!-\! \tau_{0,-}^{-1}).
\end{split}
\label{eq:wl}
\end{equation}
Here, the ultraviolet cutoff under the logarithm is, typically, $\tau_\mathrm{min}\sim\tau$, but for the description of the WL effect in graphene,~\cite{Falko2006,Savchenko2008,Savchenko2009,Falko2012,Mirlin2012,Lara2011} one should use for $\tau_\mathrm{min}$ in Eq.~\eqref{eq:wl} the intervalley scattering time $\tau_\mathrm{iv}$, instead of the mean free path time $\tau$. 
The MC, studied as a function of $B_\perp$ for fixed $B_\parallel$, takes the form 
\begin{equation}
\begin{split}
\sigma(B_\perp,B_\parallel)-\sigma(0,B_\parallel)= -\frac{e^{2}}{2\pi h} \int d\varepsilon \sum_{\alpha=\pm}
n_{F}'(\varepsilon_{\alpha}) 
\quad\\
\times \left\{
F\!\left(\!\frac{B_\perp}{B_{1,\alpha}}\!\right)
+ \left[ F\!\left(\!\frac{B_\perp}{B_{0,-}}\!\right) - F\!\left(\!\frac{B_\perp}{B_{0,+}}\!\right) \right] A_{\alpha}
\right\},  \\
F(z)=\ln z+\psi\left(\frac12+\frac{1}{z}\right), \quad B_{\beta,\alpha}=\frac{\hbar/4e}{D\tau_{\beta,\alpha}},
\end{split}
\label{eq:mc}
\end{equation}
where $\psi$ is the digamma function.

One may see that the part stemming from $C_{1/0,0}$ (square brackets) saturates at high $B_\perp$; it contributes (in units of $e^2/2\pi h$) at most $(3/4)\ln 3\approx 0.82$ to the MC.
This should be contrasted with the logarithmic growth with $B_\perp$ of the term coming from $C_{1,\pm 1}$ (the first term in braces).
That peculiarity of the field dependence sets the dynamic range of MC useful for extracting the MC curvature, $\kappa (B_\parallel)\propto \tau_\varphi^2$, using the expansion $F(z)\approx z^{2}/24 + O[z^{3}]$, as illustrated in the inset of Fig.~\ref{fig:curvatures}.

In the ``high-temperature'' limit, 
\begin{equation}
kT \gtrsim \hbar S / \tau_s |1-g_e/g_i|,
\label{eq:htlimit}
\end{equation} 
the expression for $\kappa$ can be simplified further for all values of $B_{\parallel}$, if inelastic $e$-$e$ collisions are neglected ($\tau_{T}^{-1}\rightarrow 0$),
\begin{equation}
\begin{split}
\kappa \approx \frac{e^{2}}{\pi h}
\left(\frac{D\tau_{s}}{\hbar/e}\right)^{2}
\Biggl\{\frac{2}{9}
\frac{W_{B}}{M_{1}^{2}} + \frac{4/3}{\left(1 + \frac{3}{4}\Omega_B^{2}\tau_{s}^{2}\right)^{2}}
\Biggr\},\\
W_{B} = \mathrm{ch}(2a)-\mathrm{ch}(a), \quad  a=g_i\mu_B B_\parallel/kT,  \\
M_1=\frac{(S+\frac12 )\coth[(S+\frac12 )a]-\frac12 \coth(\frac12 a)}{S (S + 1)},
\end{split}
\label{eq:csmf}
\end{equation}
where $(S+1)M_1$ is the Brillouin function [see Eq.~\eqref{eq:rse}].
The first term in braces in Eq.~\eqref{eq:csmf} comes from Cooperons $C_{1,\pm1}$ of  Eq.~\eqref{eq:WLconductivity} and has two asymptotes: $3+\frac{29+8S(1+S)}{20} (g_i\mu_BB_\parallel/kT)^2$ at $B_{\parallel}\ll kT/g_i\mu_BS$ and $\frac{1}{9}(S+1)^{2}\exp(2g_i\mu_BB_\parallel/kT)$ at $B_{\parallel}\gg kT/g_i\mu_B$. 
The latter exponential asymptote is cut off by inelastic $e$-$e$ scattering resulting in $\kappa = \frac{2e^{2}}{3\pi h} \bigl(\frac{D\tau_{T}}{\hbar/e}\bigr)^{2}$.
The second term in braces originates from $C_{1/0,0}$ and its contribution to $\kappa(B_\parallel)$ decays with increasing $B_\parallel$. Together, the two contributions provide the non-monotonic dependence of magnetoconductance curvature (which is conventionally considered as the measure of coherence time) over the in-plane field scale $B_\parallel\sim \hbar/|g_i-g_e|\mu_B\tau_s$. This nonmonotonic dependence includes a local maximum at $B_\parallel=0$ and a minimum at $B_\parallel\sim \frac{kT}{g_i\mu_BS}\bigl(\bigl|1\!-\!\frac{g_e}{g_i}\bigr|\frac{kT\tau_s}{\hbar S}\bigr)^{-2/3}$, which is followed by the increase of $\kappa (B_\parallel)$ due to the polarization of defect spins [cf.\ Eqs.~\eqref{eq:htlimit} and~\eqref{eq:csmf}].
Eventually MC curvature saturates at the scale set by the inelastic electron-electron scattering decoherence time $\tau_T$.

The above-described anomalous behavior of the decoherence rate occurs only when the electron or/and magnetic defect have $g$ factor values different from the free electron $g=-2$.
The values of $g_i \ne -2$ may be caused by the crystalline anisotropy effect on a heavy ion embedded in a 2D metal or semiconductor ({\it e.g.}, graphene).
For example, crystalline anisotropy splits states of a spin-$\frac32$ magnetic ion into two Kramers doublets with spin projections $\pm \frac12$ and $\pm \frac32$ onto the direction perpendicular to the plane of the 2D electron system.
Then, the Zeeman splitting by the in-plane magnetic field realizes the cases of $g_i=-4$ (yellow line in Fig.~\ref{fig:curvatures}) and $g_i=0$ (orange line in Fig.~\ref{fig:curvatures}) for the two doublets, respectively.%
~\footnote{%
Here, we assume that a low symmetry coordination of magnetic defect permits a generic non-diagonal form of anisotropic exchange interaction tensor, $J_{\alpha\beta}S_i^{\alpha}\sigma^{\beta}$, which provides the electron spin flips in the electron-defect scattering.%
}
The case of $g_e=-2$, $g_i\neq -2$ was realized in graphene exfoliated on a SiO$_2$/Si substrate,\cite{Folk2013} where the non-monotonic magnetoconductance is well-described\cite{Folk2015} by the theory presented here.
Alternatively, the situation $g_e=0$ (green line in Fig.~\ref{fig:curvatures}) can appear in $p$-doped transition metal dichalcogenides MoS$_2$, MoSe$_2$, WS$_2$ or WSe$_2$, where, due to a large spin-orbit splitting, Kramers doublets of the hole states correspond to opposite spins in the opposite valleys, and the external in-plane magnetic field does not lift the Kramers degeneracy.~\cite{Kormanyos2015}

In conclusion, the difference in the precession frequencies of the electron and impurity spins results in a non-monotonic dependence of the electron decoherence time on the magnetic field causing the precession.
We find the magnitude and the functional form of that dependence, and relate it to the parameters of the itinerant electrons and magnetic impurities.
Despite being small, the effect is important, as a manifestation of the very basic physics of magnetic moments in solids.

\begin{acknowledgments}
We thank I.~Aleiner, B.~Altshuler and M.~Vavilov for useful discussions.
This work has been supported by the Deutsche Forschungsgemeinschaft and Research Training Group GRK 1621, ERC Synergy Grant Hetero2D, Royal Society, and U.S.\ NSF DMR-1206612.
\end{acknowledgments}

\bibliography{wlgspin}

\begin{thebibliography}{28}%
\makeatletter
\providecommand \@ifxundefined [1]{%
 \@ifx{#1\undefined}
}%
\providecommand \@ifnum [1]{%
 \ifnum #1\expandafter \@firstoftwo
 \else \expandafter \@secondoftwo
 \fi
}%
\providecommand \@ifx [1]{%
 \ifx #1\expandafter \@firstoftwo
 \else \expandafter \@secondoftwo
 \fi
}%
\providecommand \natexlab [1]{#1}%
\providecommand \enquote  [1]{``#1''}%
\providecommand \bibnamefont  [1]{#1}%
\providecommand \bibfnamefont [1]{#1}%
\providecommand \citenamefont [1]{#1}%
\providecommand \href@noop [0]{\@secondoftwo}%
\providecommand \href [0]{\begingroup \@sanitize@url \@href}%
\providecommand \@href[1]{\@@startlink{#1}\@@href}%
\providecommand \@@href[1]{\endgroup#1\@@endlink}%
\providecommand \@sanitize@url [0]{\catcode `\\12\catcode `\$12\catcode
  `\&12\catcode `\#12\catcode `\^12\catcode `\_12\catcode `\%12\relax}%
\providecommand \@@startlink[1]{}%
\providecommand \@@endlink[0]{}%
\providecommand \url  [0]{\begingroup\@sanitize@url \@url }%
\providecommand \@url [1]{\endgroup\@href {#1}{\urlprefix }}%
\providecommand \urlprefix  [0]{URL }%
\providecommand \Eprint [0]{\href }%
\providecommand \doibase [0]{http://dx.doi.org/}%
\providecommand \selectlanguage [0]{\@gobble}%
\providecommand \bibinfo  [0]{\@secondoftwo}%
\providecommand \bibfield  [0]{\@secondoftwo}%
\providecommand \translation [1]{[#1]}%
\providecommand \BibitemOpen [0]{}%
\providecommand \bibitemStop [0]{}%
\providecommand \bibitemNoStop [0]{.\EOS\space}%
\providecommand \EOS [0]{\spacefactor3000\relax}%
\providecommand \BibitemShut  [1]{\csname bibitem#1\endcsname}%
\let\auto@bib@innerbib\@empty
\bibitem [{\citenamefont {Altshuler}\ \emph {et~al.}(1980)\citenamefont
  {Altshuler}, \citenamefont {Khmel'nitzkii}, \citenamefont {Larkin},\ and\
  \citenamefont {Lee}}]{Altshuler1980}%
  \BibitemOpen
  \bibfield  {author} {\bibinfo {author} {\bibfnamefont {B.~L.}\ \bibnamefont
  {Altshuler}}, \bibinfo {author} {\bibfnamefont {D.}~\bibnamefont
  {Khmel'nitzkii}}, \bibinfo {author} {\bibfnamefont {A.~I.}\ \bibnamefont
  {Larkin}}, \ and\ \bibinfo {author} {\bibfnamefont {P.~A.}\ \bibnamefont
  {Lee}},\ }\href {\doibase 10.1103/PhysRevB.22.5142} {\bibfield  {journal}
  {\bibinfo  {journal} {Phys. Rev. B}\ }\textbf {\bibinfo {volume} {22}},\
  \bibinfo {pages} {5142} (\bibinfo {year} {1980})}\BibitemShut {NoStop}%
\bibitem [{\citenamefont {Altshuler}\ and\ \citenamefont
  {Aronov}(1985)}]{Altshuler1985}%
  \BibitemOpen
  \bibfield  {author} {\bibinfo {author} {\bibfnamefont {B.}~\bibnamefont
  {Altshuler}}\ and\ \bibinfo {author} {\bibfnamefont {A.}~\bibnamefont
  {Aronov}},\ }in\ \href {\doibase 10.1016/B978-0-444-86916-6.50007-7} {\emph
  {\bibinfo {booktitle} {Electron--Electron Interactions in Disordered
  Systems}}},\ \bibinfo {series} {Modern Problems in Condensed Matter
  Sciences}, Vol.~\bibinfo {volume} {10},\ \bibinfo {editor} {edited by\
  \bibinfo {editor} {\bibfnamefont {A.}~\bibnamefont {Efros}}\ and\ \bibinfo
  {editor} {\bibfnamefont {M.}~\bibnamefont {Pollak}}}\ (\bibinfo  {publisher}
  {Elsevier},\ \bibinfo {address} {North-Holland, Amsterdam},\ \bibinfo {year}
  {1985})\ Chap.~\bibinfo {chapter} {1}, pp.\ \bibinfo {pages}
  {1--153}\BibitemShut {NoStop}%
\bibitem [{\citenamefont {Aleiner}\ \emph {et~al.}(1999)\citenamefont
  {Aleiner}, \citenamefont {Altshuler},\ and\ \citenamefont
  {Gershenson}}]{Aleiner1999}%
  \BibitemOpen
  \bibfield  {author} {\bibinfo {author} {\bibfnamefont {I.~L.}\ \bibnamefont
  {Aleiner}}, \bibinfo {author} {\bibfnamefont {B.~L.}\ \bibnamefont
  {Altshuler}}, \ and\ \bibinfo {author} {\bibfnamefont {M.~E.}\ \bibnamefont
  {Gershenson}},\ }\href {\doibase 10.1088/0959-7174/9/2/308} {\bibfield
  {journal} {\bibinfo  {journal} {Waves in Random Media}\ }\textbf {\bibinfo
  {volume} {9}},\ \bibinfo {pages} {201} (\bibinfo {year} {1999})}\BibitemShut
  {NoStop}%
\bibitem [{Note1()}]{Note1}%
  \BibitemOpen
  \bibinfo {note} {In fact, this contribution to the decoherence rate also
  vanishes at $T\to 0$, but the characteristic temperature scale is defined by
  the Kondo temperature~\cite {Glazman2003} and in many cases is extremely
  low.}\BibitemShut {Stop}%
\bibitem [{\citenamefont {Hikami}\ \emph {et~al.}(1980)\citenamefont {Hikami},
  \citenamefont {Larkin},\ and\ \citenamefont {Nagaoka}}]{Larkin1980}%
  \BibitemOpen
  \bibfield  {author} {\bibinfo {author} {\bibfnamefont {S.}~\bibnamefont
  {Hikami}}, \bibinfo {author} {\bibfnamefont {A.~I.}\ \bibnamefont {Larkin}},
  \ and\ \bibinfo {author} {\bibfnamefont {Y.}~\bibnamefont {Nagaoka}},\ }\href
  {\doibase 10.1143/PTP.63.707} {\bibfield  {journal} {\bibinfo  {journal}
  {Prog. Theor. Phys.}\ }\textbf {\bibinfo {volume} {63}},\ \bibinfo {pages}
  {707} (\bibinfo {year} {1980})}\BibitemShut {NoStop}%
\bibitem [{\citenamefont {Lee}\ \emph {et~al.}(1987)\citenamefont {Lee},
  \citenamefont {Stone},\ and\ \citenamefont {Fukuyama}}]{Stone1987}%
  \BibitemOpen
  \bibfield  {author} {\bibinfo {author} {\bibfnamefont {P.~A.}\ \bibnamefont
  {Lee}}, \bibinfo {author} {\bibfnamefont {A.~D.}\ \bibnamefont {Stone}}, \
  and\ \bibinfo {author} {\bibfnamefont {H.}~\bibnamefont {Fukuyama}},\ }\href
  {\doibase 10.1103/PhysRevB.35.1039} {\bibfield  {journal} {\bibinfo
  {journal} {Phys. Rev. B}\ }\textbf {\bibinfo {volume} {35}},\ \bibinfo
  {pages} {1039} (\bibinfo {year} {1987})}\BibitemShut {NoStop}%
\bibitem [{\citenamefont {Stone}(1989)}]{Stone1989}%
  \BibitemOpen
  \bibfield  {author} {\bibinfo {author} {\bibfnamefont {A.~D.}\ \bibnamefont
  {Stone}},\ }\href {\doibase 10.1103/PhysRevB.39.10736} {\bibfield  {journal}
  {\bibinfo  {journal} {Phys. Rev. B}\ }\textbf {\bibinfo {volume} {39}},\
  \bibinfo {pages} {10736} (\bibinfo {year} {1989})}\BibitemShut {NoStop}%
\bibitem [{\citenamefont {Fal'ko}(1991)}]{Falko1991}%
  \BibitemOpen
  \bibfield  {author} {\bibinfo {author} {\bibfnamefont {V.~I.}\ \bibnamefont
  {Fal'ko}},\ }\href {http://www.jetpletters.ac.ru/ps/1151/article_17419.shtml}
  {\bibfield  {journal} {\bibinfo  {journal} {JETP Lett.}\ }\textbf {\bibinfo
  {volume} {53}},\ \bibinfo {pages} {340 [Pis'ma ZhETF 53, 6, 325]} (\bibinfo
  {year} {1991})}\BibitemShut {NoStop}%
\bibitem [{\citenamefont {Chandrasekhar}\ \emph {et~al.}(1990)\citenamefont
  {Chandrasekhar}, \citenamefont {Santhanam},\ and\ \citenamefont
  {Prober}}]{Chandrasekhar1990}%
  \BibitemOpen
  \bibfield  {author} {\bibinfo {author} {\bibfnamefont {V.}~\bibnamefont
  {Chandrasekhar}}, \bibinfo {author} {\bibfnamefont {P.}~\bibnamefont
  {Santhanam}}, \ and\ \bibinfo {author} {\bibfnamefont {D.~E.}\ \bibnamefont
  {Prober}},\ }\href {\doibase 10.1103/PhysRevB.42.6823} {\bibfield  {journal}
  {\bibinfo  {journal} {Phys. Rev. B}\ }\textbf {\bibinfo {volume} {42}},\
  \bibinfo {pages} {6823} (\bibinfo {year} {1990})}\BibitemShut {NoStop}%
\bibitem [{\citenamefont {Geim}\ \emph {et~al.}(1990)\citenamefont {Geim},
  \citenamefont {Dubonos},\ and\ \citenamefont {Antonova}}]{Geim1990}%
  \BibitemOpen
  \bibfield  {author} {\bibinfo {author} {\bibfnamefont {A.~K.}\ \bibnamefont
  {Geim}}, \bibinfo {author} {\bibfnamefont {S.~V.}\ \bibnamefont {Dubonos}}, \
  and\ \bibinfo {author} {\bibfnamefont {I.~Y.}\ \bibnamefont {Antonova}},\
  }\href {http://www.jetpletters.ac.ru/ps/1155/article_17484.shtml} {\bibfield
  {journal} {\bibinfo  {journal} {JETP Lett.}\ }\textbf {\bibinfo {volume}
  {52}},\ \bibinfo {pages} {247 [Pis'ma ZhETF 52, 4, 873]} (\bibinfo {year}
  {1990})}\BibitemShut {NoStop}%
\bibitem [{\citenamefont {de~Vegvar}\ \emph {et~al.}(1991)\citenamefont
  {de~Vegvar}, \citenamefont {L\'evy},\ and\ \citenamefont
  {Fulton}}]{Levy1991}%
  \BibitemOpen
  \bibfield  {author} {\bibinfo {author} {\bibfnamefont {P.~G.~N.}\
  \bibnamefont {de~Vegvar}}, \bibinfo {author} {\bibfnamefont {L.~P.}\
  \bibnamefont {L\'evy}}, \ and\ \bibinfo {author} {\bibfnamefont {T.~A.}\
  \bibnamefont {Fulton}},\ }\href {\doibase 10.1103/PhysRevLett.66.2380}
  {\bibfield  {journal} {\bibinfo  {journal} {Phys. Rev. Lett.}\ }\textbf
  {\bibinfo {volume} {66}},\ \bibinfo {pages} {2380} (\bibinfo {year}
  {1991})}\BibitemShut {NoStop}%
\bibitem [{\citenamefont {Gougam}\ \emph {et~al.}(2000)\citenamefont {Gougam},
  \citenamefont {Pierre}, \citenamefont {Pothier}, \citenamefont {Esteve},\
  and\ \citenamefont {Birge}}]{Birge2000}%
  \BibitemOpen
  \bibfield  {author} {\bibinfo {author} {\bibfnamefont {A.~B.}\ \bibnamefont
  {Gougam}}, \bibinfo {author} {\bibfnamefont {F.}~\bibnamefont {Pierre}},
  \bibinfo {author} {\bibfnamefont {H.}~\bibnamefont {Pothier}}, \bibinfo
  {author} {\bibfnamefont {D.}~\bibnamefont {Esteve}}, \ and\ \bibinfo {author}
  {\bibfnamefont {N.~O.}\ \bibnamefont {Birge}},\ }\href {\doibase
  10.1023/A:1004658404535} {\bibfield  {journal} {\bibinfo  {journal} {J. Low
  Temp. Phys.}\ }\textbf {\bibinfo {volume} {118}},\ \bibinfo {pages} {447}
  (\bibinfo {year} {2000})}\BibitemShut {NoStop}%
\bibitem [{\citenamefont {Pierre}\ and\ \citenamefont
  {Birge}(2002)}]{Birge2002}%
  \BibitemOpen
  \bibfield  {author} {\bibinfo {author} {\bibfnamefont {F.}~\bibnamefont
  {Pierre}}\ and\ \bibinfo {author} {\bibfnamefont {N.~O.}\ \bibnamefont
  {Birge}},\ }\href {\doibase 10.1103/PhysRevLett.89.206804} {\bibfield
  {journal} {\bibinfo  {journal} {Phys. Rev. Lett.}\ }\textbf {\bibinfo
  {volume} {89}},\ \bibinfo {pages} {206804} (\bibinfo {year}
  {2002})}\BibitemShut {NoStop}%
\bibitem [{\citenamefont {Pierre}\ \emph {et~al.}(2003)\citenamefont {Pierre},
  \citenamefont {Gougam}, \citenamefont {Anthore}, \citenamefont {Pothier},
  \citenamefont {Esteve},\ and\ \citenamefont {Birge}}]{Birge2003}%
  \BibitemOpen
  \bibfield  {author} {\bibinfo {author} {\bibfnamefont {F.}~\bibnamefont
  {Pierre}}, \bibinfo {author} {\bibfnamefont {A.~B.}\ \bibnamefont {Gougam}},
  \bibinfo {author} {\bibfnamefont {A.}~\bibnamefont {Anthore}}, \bibinfo
  {author} {\bibfnamefont {H.}~\bibnamefont {Pothier}}, \bibinfo {author}
  {\bibfnamefont {D.}~\bibnamefont {Esteve}}, \ and\ \bibinfo {author}
  {\bibfnamefont {N.~O.}\ \bibnamefont {Birge}},\ }\href {\doibase
  10.1103/PhysRevB.68.085413} {\bibfield  {journal} {\bibinfo  {journal} {Phys.
  Rev. B}\ }\textbf {\bibinfo {volume} {68}},\ \bibinfo {pages} {085413}
  (\bibinfo {year} {2003})}\BibitemShut {NoStop}%
\bibitem [{\citenamefont {Vavilov}\ and\ \citenamefont
  {Glazman}(2003)}]{Glazman2003}%
  \BibitemOpen
  \bibfield  {author} {\bibinfo {author} {\bibfnamefont {M.~G.}\ \bibnamefont
  {Vavilov}}\ and\ \bibinfo {author} {\bibfnamefont {L.~I.}\ \bibnamefont
  {Glazman}},\ }\href {\doibase 10.1103/PhysRevB.67.115310} {\bibfield
  {journal} {\bibinfo  {journal} {Phys. Rev. B}\ }\textbf {\bibinfo {volume}
  {67}},\ \bibinfo {pages} {115310} (\bibinfo {year} {2003})}\BibitemShut
  {NoStop}%
\bibitem [{\citenamefont {Fal'ko}(1992)}]{Falko1992}%
  \BibitemOpen
  \bibfield  {author} {\bibinfo {author} {\bibfnamefont {V.~I.}\ \bibnamefont
  {Fal'ko}},\ }\href {\doibase 10.1088/0953-8984/4/15/009} {\bibfield
  {journal} {\bibinfo  {journal} {J. Phys.: Condens. Matter}\ }\textbf
  {\bibinfo {volume} {4}},\ \bibinfo {pages} {3943} (\bibinfo {year}
  {1992})}\BibitemShut {NoStop}%
\bibitem [{\citenamefont {Amaral}(1990)}]{Amaral1990}%
  \BibitemOpen
  \bibfield  {author} {\bibinfo {author} {\bibfnamefont {V.~S.}\ \bibnamefont
  {Amaral}},\ }\href {\doibase 10.1088/0953-8984/2/41/008} {\bibfield
  {journal} {\bibinfo  {journal} {J. Phys.: Condens. Matter}\ }\textbf
  {\bibinfo {volume} {2}},\ \bibinfo {pages} {8201} (\bibinfo {year}
  {1990})}\BibitemShut {NoStop}%
\bibitem [{Note2()}]{Note2}%
  \BibitemOpen
  \bibinfo {note} {Note $g_{i}$ (not $g_{e}$) in this equation, which is
  because the Green's function of an electron with spin $\sigma $ has energy
  $\varepsilon _{\sigma }$. The energy of the electron changes only at the spin
  flip, always by the value $\pm g_{i}\mu _{B}B_{\parallel }$ dependent on the
  sign of the transmitted spin $\sigma -\sigma '$. Therefore, in the final spin
  state $\sigma '$, the electron will have the energy $\varepsilon _{\sigma
  }\pm g_{i}\mu _{B}B_{\parallel }=\varepsilon _{\sigma '}=\varepsilon
  _{-\sigma }$.}\BibitemShut {Stop}%
\bibitem [{\citenamefont {McCann}\ \emph {et~al.}(2006)\citenamefont {McCann},
  \citenamefont {Kechedzhi}, \citenamefont {Fal'ko}, \citenamefont {Suzuura},
  \citenamefont {Ando},\ and\ \citenamefont {Altshuler}}]{Falko2006}%
  \BibitemOpen
  \bibfield  {author} {\bibinfo {author} {\bibfnamefont {E.}~\bibnamefont
  {McCann}}, \bibinfo {author} {\bibfnamefont {K.}~\bibnamefont {Kechedzhi}},
  \bibinfo {author} {\bibfnamefont {V.~I.}\ \bibnamefont {Fal'ko}}, \bibinfo
  {author} {\bibfnamefont {H.}~\bibnamefont {Suzuura}}, \bibinfo {author}
  {\bibfnamefont {T.}~\bibnamefont {Ando}}, \ and\ \bibinfo {author}
  {\bibfnamefont {B.~L.}\ \bibnamefont {Altshuler}},\ }\href {\doibase
  10.1103/PhysRevLett.97.146805} {\bibfield  {journal} {\bibinfo  {journal}
  {Phys. Rev. Lett.}\ }\textbf {\bibinfo {volume} {97}},\ \bibinfo {pages}
  {146805} (\bibinfo {year} {2006})}\BibitemShut {NoStop}%
\bibitem [{\citenamefont {Tikhonenko}\ \emph {et~al.}(2008)\citenamefont
  {Tikhonenko}, \citenamefont {Horsell}, \citenamefont {Gorbachev},\ and\
  \citenamefont {Savchenko}}]{Savchenko2008}%
  \BibitemOpen
  \bibfield  {author} {\bibinfo {author} {\bibfnamefont {F.~V.}\ \bibnamefont
  {Tikhonenko}}, \bibinfo {author} {\bibfnamefont {D.~W.}\ \bibnamefont
  {Horsell}}, \bibinfo {author} {\bibfnamefont {R.~V.}\ \bibnamefont
  {Gorbachev}}, \ and\ \bibinfo {author} {\bibfnamefont {A.~K.}\ \bibnamefont
  {Savchenko}},\ }\href {\doibase 10.1103/PhysRevLett.100.056802} {\bibfield
  {journal} {\bibinfo  {journal} {Phys. Rev. Lett.}\ }\textbf {\bibinfo
  {volume} {100}},\ \bibinfo {pages} {056802} (\bibinfo {year}
  {2008})}\BibitemShut {NoStop}%
\bibitem [{\citenamefont {Tikhonenko}\ \emph {et~al.}(2009)\citenamefont
  {Tikhonenko}, \citenamefont {Kozikov}, \citenamefont {Savchenko},\ and\
  \citenamefont {Gorbachev}}]{Savchenko2009}%
  \BibitemOpen
  \bibfield  {author} {\bibinfo {author} {\bibfnamefont {F.~V.}\ \bibnamefont
  {Tikhonenko}}, \bibinfo {author} {\bibfnamefont {A.~A.}\ \bibnamefont
  {Kozikov}}, \bibinfo {author} {\bibfnamefont {A.~K.}\ \bibnamefont
  {Savchenko}}, \ and\ \bibinfo {author} {\bibfnamefont {R.~V.}\ \bibnamefont
  {Gorbachev}},\ }\href {\doibase 10.1103/PhysRevLett.103.226801} {\bibfield
  {journal} {\bibinfo  {journal} {Phys. Rev. Lett.}\ }\textbf {\bibinfo
  {volume} {103}},\ \bibinfo {pages} {226801} (\bibinfo {year}
  {2009})}\BibitemShut {NoStop}%
\bibitem [{\citenamefont {Kozikov}\ \emph {et~al.}(2012)\citenamefont
  {Kozikov}, \citenamefont {Horsell}, \citenamefont {McCann},\ and\
  \citenamefont {Fal'ko}}]{Falko2012}%
  \BibitemOpen
  \bibfield  {author} {\bibinfo {author} {\bibfnamefont {A.~A.}\ \bibnamefont
  {Kozikov}}, \bibinfo {author} {\bibfnamefont {D.~W.}\ \bibnamefont
  {Horsell}}, \bibinfo {author} {\bibfnamefont {E.}~\bibnamefont {McCann}}, \
  and\ \bibinfo {author} {\bibfnamefont {V.~I.}\ \bibnamefont {Fal'ko}},\
  }\href {\doibase 10.1103/PhysRevB.86.045436} {\bibfield  {journal} {\bibinfo
  {journal} {Phys. Rev. B}\ }\textbf {\bibinfo {volume} {86}},\ \bibinfo
  {pages} {045436} (\bibinfo {year} {2012})}\BibitemShut {NoStop}%
\bibitem [{\citenamefont {Jobst}\ \emph {et~al.}(2012)\citenamefont {Jobst},
  \citenamefont {Waldmann}, \citenamefont {Gornyi}, \citenamefont {Mirlin},\
  and\ \citenamefont {Weber}}]{Mirlin2012}%
  \BibitemOpen
  \bibfield  {author} {\bibinfo {author} {\bibfnamefont {J.}~\bibnamefont
  {Jobst}}, \bibinfo {author} {\bibfnamefont {D.}~\bibnamefont {Waldmann}},
  \bibinfo {author} {\bibfnamefont {I.~V.}\ \bibnamefont {Gornyi}}, \bibinfo
  {author} {\bibfnamefont {A.~D.}\ \bibnamefont {Mirlin}}, \ and\ \bibinfo
  {author} {\bibfnamefont {H.~B.}\ \bibnamefont {Weber}},\ }\href {\doibase
  10.1103/PhysRevLett.108.106601} {\bibfield  {journal} {\bibinfo  {journal}
  {Phys. Rev. Lett.}\ }\textbf {\bibinfo {volume} {108}},\ \bibinfo {pages}
  {106601} (\bibinfo {year} {2012})}\BibitemShut {NoStop}%
\bibitem [{\citenamefont {Lara-Avila}\ \emph {et~al.}(2011)\citenamefont
  {Lara-Avila}, \citenamefont {Tzalenchuk}, \citenamefont {Kubatkin},
  \citenamefont {Yakimova}, \citenamefont {Janssen}, \citenamefont {Cedergren},
  \citenamefont {Bergsten},\ and\ \citenamefont {Fal'ko}}]{Lara2011}%
  \BibitemOpen
  \bibfield  {author} {\bibinfo {author} {\bibfnamefont {S.}~\bibnamefont
  {Lara-Avila}}, \bibinfo {author} {\bibfnamefont {A.}~\bibnamefont
  {Tzalenchuk}}, \bibinfo {author} {\bibfnamefont {S.}~\bibnamefont
  {Kubatkin}}, \bibinfo {author} {\bibfnamefont {R.}~\bibnamefont {Yakimova}},
  \bibinfo {author} {\bibfnamefont {T.~J. B.~M.}\ \bibnamefont {Janssen}},
  \bibinfo {author} {\bibfnamefont {K.}~\bibnamefont {Cedergren}}, \bibinfo
  {author} {\bibfnamefont {T.}~\bibnamefont {Bergsten}}, \ and\ \bibinfo
  {author} {\bibfnamefont {V.}~\bibnamefont {Fal'ko}},\ }\href {\doibase
  10.1103/PhysRevLett.107.166602} {\bibfield  {journal} {\bibinfo  {journal}
  {Phys. Rev. Lett.}\ }\textbf {\bibinfo {volume} {107}},\ \bibinfo {pages}
  {166602} (\bibinfo {year} {2011})}\BibitemShut {NoStop}%
\bibitem [{Note3()}]{Note3}%
  \BibitemOpen
  \bibinfo {note} {Here, we assume that a low symmetry coordination of magnetic
  defect permits a generic non-diagonal form of anisotropic exchange
  interaction tensor, $J_{\alpha \beta }S_i^{\alpha }\sigma ^{\beta }$, which
  provides the electron spin flips in the electron-defect
  scattering.}\BibitemShut {Stop}%
\bibitem [{\citenamefont {Lundeberg}\ \emph {et~al.}(2013)\citenamefont
  {Lundeberg}, \citenamefont {Yang}, \citenamefont {Renard},\ and\
  \citenamefont {Folk}}]{Folk2013}%
  \BibitemOpen
  \bibfield  {author} {\bibinfo {author} {\bibfnamefont {M.~B.}\ \bibnamefont
  {Lundeberg}}, \bibinfo {author} {\bibfnamefont {R.}~\bibnamefont {Yang}},
  \bibinfo {author} {\bibfnamefont {J.}~\bibnamefont {Renard}}, \ and\ \bibinfo
  {author} {\bibfnamefont {J.~A.}\ \bibnamefont {Folk}},\ }\href {\doibase
  10.1103/PhysRevLett.110.156601} {\bibfield  {journal} {\bibinfo  {journal}
  {Phys. Rev. Lett.}\ }\textbf {\bibinfo {volume} {110}},\ \bibinfo {pages}
  {156601} (\bibinfo {year} {2013})}\BibitemShut {NoStop}%
\bibitem [{\citenamefont {Lara-Avila}\ \emph {et~al.}(2015)\citenamefont
  {Lara-Avila}, \citenamefont {Kubatkin}, \citenamefont {Kashuba},
  \citenamefont {Folk}, \citenamefont {L\"uscher}, \citenamefont {Yakimova},
  \citenamefont {Janssen}, \citenamefont {Tzalenchuk},\ and\ \citenamefont
  {Fal'ko}}]{Folk2015}%
  \BibitemOpen
  \bibfield  {author} {\bibinfo {author} {\bibfnamefont {S.}~\bibnamefont
  {Lara-Avila}}, \bibinfo {author} {\bibfnamefont {S.}~\bibnamefont
  {Kubatkin}}, \bibinfo {author} {\bibfnamefont {O.}~\bibnamefont {Kashuba}},
  \bibinfo {author} {\bibfnamefont {J.~A.}\ \bibnamefont {Folk}}, \bibinfo
  {author} {\bibfnamefont {S.}~\bibnamefont {L\"uscher}}, \bibinfo {author}
  {\bibfnamefont {R.}~\bibnamefont {Yakimova}}, \bibinfo {author}
  {\bibfnamefont {T.~J. B.~M.}\ \bibnamefont {Janssen}}, \bibinfo {author}
  {\bibfnamefont {A.}~\bibnamefont {Tzalenchuk}}, \ and\ \bibinfo {author}
  {\bibfnamefont {V.}~\bibnamefont {Fal'ko}},\ }\href {\doibase
  10.1103/PhysRevLett.115.106602} {\bibfield  {journal} {\bibinfo  {journal}
  {Phys. Rev. Lett.}\ }\textbf {\bibinfo {volume} {115}},\ \bibinfo {pages}
  {106602} (\bibinfo {year} {2015})}\BibitemShut {NoStop}%
\bibitem [{\citenamefont {Korm\'anyos}\ \emph {et~al.}(2015)\citenamefont
  {Korm\'anyos}, \citenamefont {Burkard}, \citenamefont {Gmitra}, \citenamefont
  {Fabian}, \citenamefont {Z\'olyomi}, \citenamefont {Drummond},\ and\
  \citenamefont {Fal'ko}}]{Kormanyos2015}%
  \BibitemOpen
  \bibfield  {author} {\bibinfo {author} {\bibfnamefont {A.}~\bibnamefont
  {Korm\'anyos}}, \bibinfo {author} {\bibfnamefont {G.}~\bibnamefont
  {Burkard}}, \bibinfo {author} {\bibfnamefont {M.}~\bibnamefont {Gmitra}},
  \bibinfo {author} {\bibfnamefont {J.}~\bibnamefont {Fabian}}, \bibinfo
  {author} {\bibfnamefont {V.}~\bibnamefont {Z\'olyomi}}, \bibinfo {author}
  {\bibfnamefont {N.~D.}\ \bibnamefont {Drummond}}, \ and\ \bibinfo {author}
  {\bibfnamefont {V.}~\bibnamefont {Fal'ko}},\ }\href {\doibase
  10.1088/2053-1583/2/2/022001} {\bibfield  {journal} {\bibinfo  {journal} {2D
  Materials}\ }\textbf {\bibinfo {volume} {2}},\ \bibinfo {pages} {022001}
  (\bibinfo {year} {2015})}\BibitemShut {NoStop}%
\end{thebibliography}%

\end{document}